\documentclass{ws-ppl}
\usepackage{tikz,pgfplots,pgfplotstable,booktabs,array,dcolumn,url}
\usepgfplotslibrary{groupplots}
\pgfplotstableset{create on use/grate/.style={create col/expr={\thisrow{rate}/1000.0}}}
\newcommand{\aplot}[6]{%
  \addplot+[#6] table[x = N, y = #1, col sep = comma] {results/#2-#3-#4.txt};
  \addlegendentry{#5}
}
\newcommand{\boltplot}[3]{\aplot{#1}{#2}{#3}{bolt}{Bolt}{color = red, mark = star}}
\newcommand{\clogsplot}[3]{\aplot{#1}{#2}{#3}{clogs}{CLOGS}{color = blue, mark = triangle}}
\newcommand{\computeplot}[3]{\aplot{#1}{#2}{#3}{compute}{Compute}{color = brown, mark = +}}
\newcommand{\cubplot}[3]{\aplot{#1}{#2}{#3}{cub}{CUB}{color = green, mark = oplus}}
\newcommand{\thrustplot}[3]{\aplot{#1}{#2}{#3}{thrust}{Thrust}{color = green!50!black, mark = square}}
\newcommand{\vexplot}[3]{\aplot{#1}{#2}{#3}{vex}{VexCL}{color = purple, mark = o}}
\newcommand{\mgpuplot}[3]{\aplot{#1}{#2}{#3}{mgpu}{MGPU}{color = gray, solid, mark = diamond}}
\newcommand{\nvplots}[3]{
  \clogsplot{#1}{#2}{#3}
  \computeplot{#1}{#2}{#3}
  \vexplot{#1}{#2}{#3}
  \thrustplot{#1}{#2}{#3}
  \cubplot{#1}{#2}{#3}
  \mgpuplot{#1}{#2}{#3}
}
\newcommand{\amdplots}[3]{
  \clogsplot{#1}{#2}{#3}
  \computeplot{#1}{#2}{#3}
  \vexplot{#1}{#2}{#3}
  \boltplot{#1}{#2}{#3}
}
\newcommand{\allplots}[3]{%
  \begin{tikzpicture}
    \begin{groupplot}[
        group style = {
          group size = 2 by 2,
          vertical sep = 2cm,
          ylabels at = {edge left}
        },
        xmode = log,
        height = 6cm,
        grid = major,
        xlabel = {Elements},
        legend style = {
          font = \tiny,
          row sep = -2.5pt,
          at = {(0.02, 0.98)},
          anchor = {north west}
        },
        #3
      ]
      \nextgroupplot[title = {480 GTX}]
      \nvplots{#1}{480gtx}{#2}

      \nextgroupplot[title = {K40}]
      \nvplots{#1}{k40c}{#2}

      \nextgroupplot[title = {R9~270}]
      \amdplots{#1}{r9-270}{#2}
    \end{groupplot}
  \end{tikzpicture}
}
\tikzset{every mark/.append style={scale = 0.5}}

\begin{document}

\markboth{Parallel Processing Letters}
{A Performance Comparison of Sort and Scan Libraries for GPUs}

%
\catchline{}{}{}{}{}
%

\title{A PERFORMANCE COMPARISON OF SORT AND SCAN LIBRARIES FOR GPUS}

\author{BRUCE MERRY\footnote{bmerry@gmail.com}}

\address{Department of Computer Science, University of Cape Town, South
Africa}

\maketitle

\begin{history}
\received{(received date)}
\revised{(revised date)}
\comby{(Name of Editor)}
\end{history}

\begin{abstract}
Sorting and scanning are two fundamental primitives for constructing highly
parallel algorithms. A number of libraries now provide implementations of
these primitives for GPUs, but there is relatively little information about
the performance of these implementations.

We benchmark seven libraries for 32-bit integer scan and sort, and sorting
32-bit values by 32-bit integer keys. We show that there is a large
variation in performance between the libraries, and that no one library has
both optimal performance and portability.
\end{abstract}


\section{Introduction}
While GPU programming languages like CUDA \cite{cuda} and OpenCL \cite{opencl} make it easy to write
code for embarrassingly parallel problems in which all elements are
independent, it is less easy to solve problems requiring \emph{cooperative}
parallelism. One of the most important primitives for such problems is the
\emph{scan}, or parallel prefix sum \cite{blelloch-scan}. Sorting is a
fundamental tool in algorithm design, and has been successfully applied in GPU
computing \cite{owens-gpgpu}.

There are now a large number of libraries that have been written on top of
OpenCL and CUDA to provide scan, sorting and other parallel primitives. However, these
primitives are challenging to optimize, and the implementations vary widely in
performance. We provide benchmark results of these libraries so that users can make
informed decisions about whether a particular library will meet their needs,
or will become a bottleneck.

We have chosen three specific problems: scanning 32-bit integers, sorting
32-bit unsigned integers, and sorting 32-bit values by 32-bit unsigned integer
keys. For current GPU memory sizes, 32 bits is enough to represent
an array index, which is why we use this size. Note that when sorting large
objects, it is more bandwidth-efficient to first sort 32-bit indices into the
array and then apply the resulting permutation.

\section{Setup}
In our experiment we have tested seven libraries:
CLOGS 1.3.0\footnote{\url{http://clogs.sourceforge.net}},
VexCL 1.3.1\footnote{\url{https://github.com/ddemidov/vexcl}} and
Boost.Compute 0.2\footnote{\url{https://github.com/kylelutz/compute}; it is
not currently an official Boost library} use OpenCL;
Bolt 1.1 also uses OpenCL, but relies on
AMD-specific extensions; and Thrust 1.7.1\footnote{Shipped as part of CUDA
6.0}, CUB 1.3.1\footnote{\url{http://nvlabs.github.io/cub/}} and
Modern GPU\footnote{\url{http://nvlabs.github.io/moderngpu/}, revision
\texttt{b6b3ed5}} use NVIDIA's CUDA API.

CLOGS is a small library that implements only sort and scan. It relies on
auto-tuning to achieve good results across a range of GPUs, but it is somewhat
rigid: it is only possible to sort or scan a small set of built-in types, and
it is not possible to provide custom addition or comparison operators.

VexCL and Boost.Compute are C++ template libraries that provide more
flexibility than CLOGS, synthesizing OpenCL kernels based on arguments
provided by the user. VexCL also provides a choice
of CUDA and OpenCL backends, but we only benchmark the OpenCL backend as a
previous benchmark of VexCL found little difference between the backends
for large problem sizes \cite{opencl-libraries}.

Thrust is an STL-like library which has backends for CUDA and for multi-core
CPUs. CUB and Modern GPU are more closely tied to CUDA, and do not hide it
under an abstraction layer. Modern GPU is intended to be easier to read and
modify, while CUB aims for maximum performance \cite{cub}.

We use three GPUs in our benchmarks:
an AMD Radeon R9~270,
an NVIDIA GeForce GTX 480
and an NVIDIA Tesla K40.
The first two are desktop GPUs while
the K40 is a server GPU. The R9~270 is based
on the GCN architecture, the GTX 480 on the older Fermi architecture, and the
K40 on the newer Kepler architecture. Table~\ref{tbl:devices} summarizes the theoretical
performance characteristics of each device.
\begin{table}
  \centering
  \caption{GPUs used in the benchmarks. The Tesla K40 was run with ECC
  disabled.}
  \label{tbl:devices}
  \footnotesize Table~\ref{tbl:devices}. GPUs used in the benchmarks. The Tesla K40 was run with ECC
  disabled.

  \vspace{1ex}\noindent
  \begin{tabular}{lD{.}{.}{-1}D{.}{.}{-1}D{.}{.}{-1}}
    \toprule
    \textbf{Device} &
      \multicolumn{1}{c}{\textbf{RAM}} &
      \multicolumn{1}{c}{\textbf{Bandwidth}} &
      \multicolumn{1}{c}{\textbf{Single precision}}\\
      &
      \multicolumn{1}{c}{\textbf{(GiB)}} &
      \multicolumn{1}{c}{\textbf{(GiB/s)}} &
      \multicolumn{1}{c}{\textbf{(TFLOP/s)}}\\
    \midrule
    Radeon R9~270 & 2 & 179.2 & 2.3\\
    GeForce 480 GTX & 1.5 & 177.4 & 1.345\\
    Tesla K40 & 12 & 288 & 4.29\\
    \bottomrule
  \end{tabular}
\end{table}

For each benchmark and library, we use the same approach. After allocating and
populating buffers, we perform ten iterations as a warmup pass, and synchronize with
the GPU. We then perform 50 iterations of the algorithm before again
synchronizing with the GPU. The runtime is measured between the two
synchronization points. We thus do \emph{not} measure the time for data
transfer between the host and GPU; we assume that in practical use the sort
or scan forms part of a larger GPU algorithm.

For the sorting algorithms, we
start by copying the data from pre-generated random buffers, and this copy
time is included in the measurements; but since the sort is far more expensive
than the initial copy, this does not affect results by more than a few
percent. Where possible, we also allocate any scratch buffers required by the
algorithm outside the loop. Currently only CLOGS and CUB support this.

We have used a range of problem sizes, consisting of powers of two, and 1--9
times a power of ten, ranging from $10^4$ to $10^8$. GPUs are poorly suited to
small problem sizes, and we consider the results for the larger sizes to be more
interesting; nevertheless, we include the results for the smaller sizes since
it shows how well the different libraries cope with limited parallelism. The
power-of-two sizes are included because they expose some unexpected
sensitivities to problem size.

\section{Results}
\subsection{Scan}
\begin{figure}[hb]
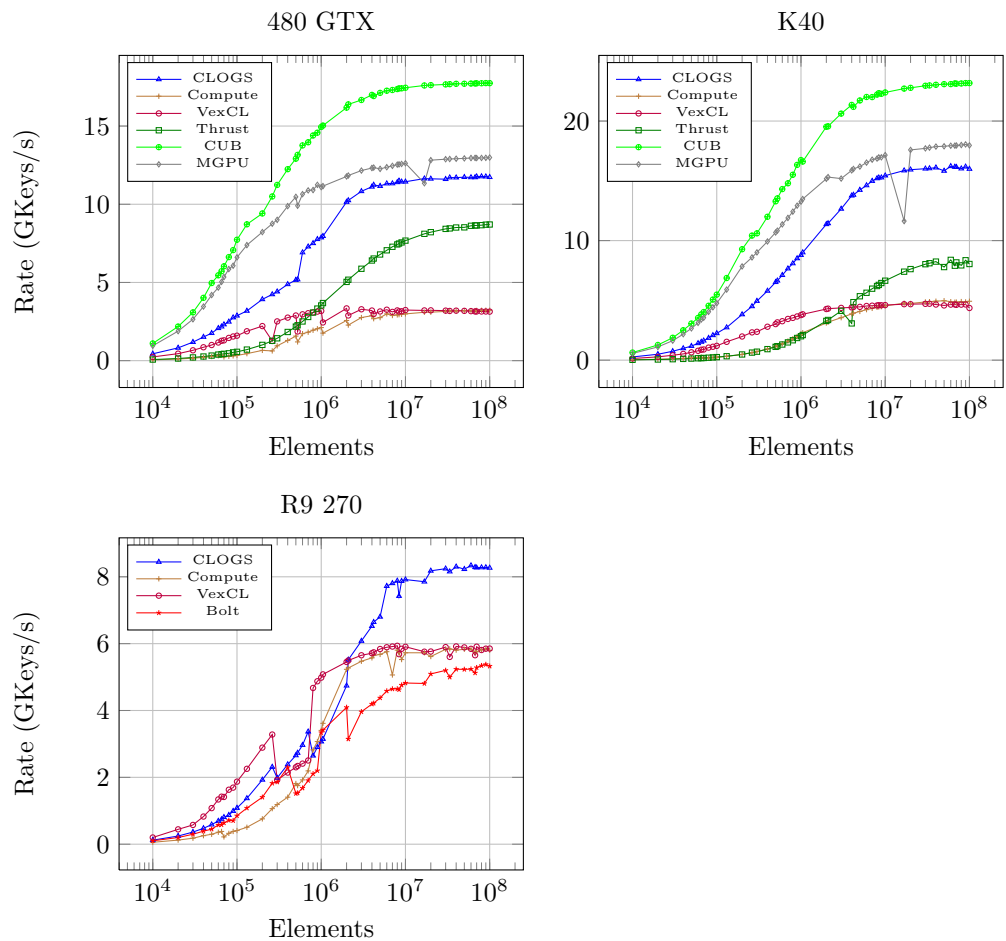

  \[\allplots{grate}{scan}{ylabel = {Rate (GKeys/s)}}\]
  \caption{Scan performance for 32-bit elements}\label{fig:scan}
\end{figure}

Figure~\ref{fig:scan} shows the results for scanning 32-bit elements. On the R9~270,
CLOGS achieves the highest throughput, while VexCL performs the best for
problem sizes under two million elements. Bolt's performance is surprisingly
poor, given that it is provided by AMD and so might be expected to be
well-tuned for AMD hardware.

On the 480 GTX, CUB has the highest performance, followed by Modern GPU. CLOGS
is the fastest of the OpenCL libraries, and outperforms NVIDIA's Thrust
library. For the largest problem sizes, VexCL and Boost.Compute achieve less
than 20\% of the performance of CUB. The picture for the K40 is essentially
the same, except that the performance drop for Modern GPU at $2^{24}$ elements
is more pronounced.

CUB's performance is notable because a simple bandwidth calculation shows that
it must perform fewer than three DRAM accesses per element. It seems unlikely
that any implementation that depends only the guarantees of OpenCL 1.2 (which
does not provide a method for inter-workgroup communication) would be able to
match this.

\subsection{Sort}
\begin{figure}[hb]
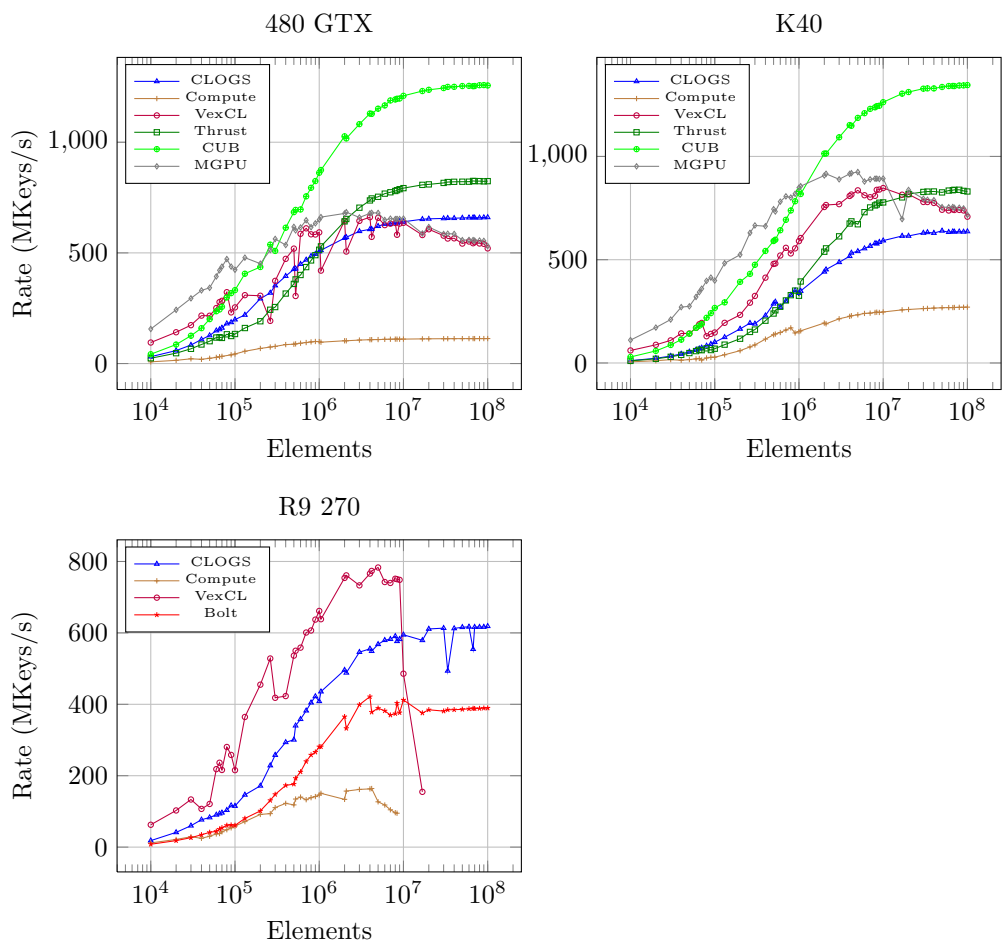

  \[\allplots{rate}{sort}{ylabel = {Rate (MKeys/s)}}\]
  \caption{Sort performance for 32-bit unsigned integers}\label{fig:sort}
\end{figure}

Figure~\ref{fig:sort} shows the results for sorting 32-bit unsigned integer
keys. On the R9~270, CLOGS and Bolt are
able to handle problem sizes all the way up to 100 million elements, while
VexCL and Boost.Compute suffer poor performance or run out of memory on much
smaller problems. Surprisingly, the 480 GTX has less
memory, yet does not display these effects. The cause may thus be poor memory
management in the driver rather than excessive memory use in the library.
VexCL has the best performance on problem sizes that it handles, while CLOGS
does best on 10 million or more elements. As for scan, the AMD-specific Bolt
is not able to keep up with generic libraries.

On the NVIDIA GPUs, CUB is again the fastest library for large problems, but
this time Modern GPU does better on small problem sizes. One difference between
the two GPUs is that CLOGS eventually surpasses Modern GPU and VexCL on the 480
GTX, but not on the K40. In fact, CLOGS sorts fewer keys per second on the K40
than the 480 GTX, even though the K40 has better theoretical performance, and
in spite of the autotuning support in CLOGS. This suggests that tuning
Fermi-optimized code for Kepler may require more than just tweaking a few
tuning parameters.

While the other libraries all achieve acceptable performance (generally at
least 40\% of the fastest), Boost.Compute performs poorly on all three GPUs,
and in some cases achieves only 10\% of the best performance.

\subsection{Sort by key}
\begin{figure}[hb]
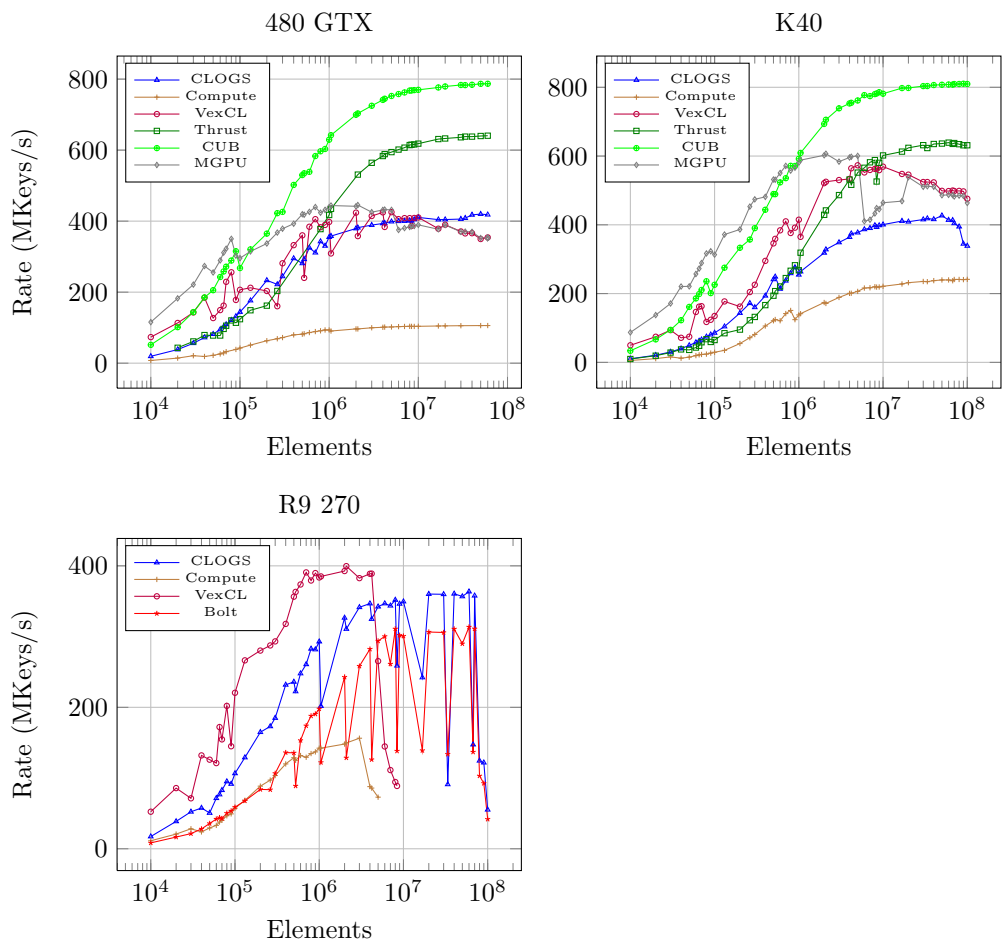

  \[\allplots{rate}{sort-by-key}{ylabel = {Rate (MKeys/s)}}\]
  \caption{Sort by Key performance for 32-bit unsigned integer keys and values}\label{fig:sort-by-key}
\end{figure}

Figure~\ref{fig:sort-by-key} shows the results for sorting 32-bit values by
32-bit keys. The results for the R9~270 show a number of interesting phenomena.
Firstly, VexCL and Boost.Compute show the same pattern of scaling up to a
certain point, beyond which performance degrades drastically, followed by
out-of-memory errors. While less obvious, the other two libraries also
show this falloff beyond 70 million elements. Beyond this point, the total
memory allocated exceeds the memory capacity of the GPU (since we need the
unordered data, the sorted data, and a scratch buffer for the keys and the
values), and the driver is presumably swapping out GPU memory to allow the
kernels to run.

Secondly, the performance of CLOGS and Bolt is highly sensitive to the problem
size. Sizes that are powers of two reduce the throughput massively, in some
cases by 75\%. We do not know why these problem sizes should cause such poor
performance.

On the NVIDIA GPUs, the situation is generally similar to sorting just keys:
Modern GPU is fastest for small sizes, CUB for larges sizes, and CLOGS does worse
on the K40 than the 480 GTX. One difference is that Modern GPU loses performance
between 6 million and 16 million elements on the K40. Also note that the plots
for the 480 GTX stop at 60 million elements, at which point the GPU memory is
exhausted.

\section{Conclusions}
The performance results show a surprising amount of variation, and no single
library provides optimal performance across multiple devices and problem
sizes. For CUDA applications, CUB has significantly higher performance than
any of the alternatives, except for small sorts; for OpenCL applications,
CLOGS gives reasonable all-round performance, while VexCL does well at sorting.
If one is only interested in sorting and scanning rather than other features
of the library, then Thrust, Bolt and Boost.Compute do not seem to offer any
advantages.

Of course, one should not choose a parallel programming library based only on
the performance of these primitives. Features, ease-of-use, portability,
performance in other areas, and interoperability with other libraries are also
important considerations.
We recommend that library designers provide low-level access to the data
structures they use, so that users can more easily mix and match libraries to
choose the best one for each primitive. As an example, VexCL allows one to
obtain the OpenCL memory objects encapsulated by a \texttt{vex::vector}, and
it provides example wrappers to perform sorting and scanning through CLOGS or
Boost.Compute.

\section*{Acknowledgements}
Denis Demidov kindly provided access to the Tesla K40 used in the experiments.
The author received funding from the South African Centre for High Performance
Computing.

\bibliography{ssbench}

\begin{thebibliography}{10}

\bibitem{mgpu-mergesort}
Sean Baxter.
\newblock {\em Modern {GPU} --- Mergesort}.
\newblock \url{http://nvlabs.github.io/moderngpu/mergesort.html}. Retrieved
  2014-08-23.

\bibitem{blelloch-scan}
G.E. Blelloch.
\newblock Scans as primitive parallel operations.
\newblock {\em Computers, IEEE Transactions on}, 38(11):1526--1538, Nov 1989.

\bibitem{opencl-libraries}
D.~Demidov, K.~Ahnert, K.~Rupp, and P.~Gottschling.
\newblock Programming {CUDA} and {OpenCL}: A case study using modern {C++}
  libraries.
\newblock {\em SIAM Journal on Scientific Computing}, 35(5):C453--C472, 2013.

\bibitem{gpu-mergepath}
Oded Green, Robert McColl, and David~A. Bader.
\newblock {GPU} merge path: A {GPU} merging algorithm.
\newblock In {\em Proceedings of the 26th ACM International Conference on
  Supercomputing}, ICS '12, pages 331--340, New York, NY, USA, 2012. ACM.

\bibitem{opencl}
{Khronos OpenCL Working Group}.
\newblock The {OpenCL} specification version 1.2, November 2011.
\newblock \url{http://www.khronos.org/registry/cl/specs/opencl-1.2.pdf}.

\bibitem{cub}
Duane Merrill.
\newblock {\em {CUB} Documentation}.
\newblock \url{http://nvlabs.github.io/cub}. Retrieved 2014-06-18.

\bibitem{merrill-scan}
Duane Merrill and Andrew Grimshaw.
\newblock Parallel scan for stream architectures.
\newblock Technical Report CS2009-14, Department of Computer Science,
  University of Virgina, Dec 2009.

\bibitem{merrill-sort}
Duane Merrill and Andrew Grimshaw.
\newblock Revisiting sorting for {GPGPU} stream architectures.
\newblock (CS2010-03), Feb 2010.

\bibitem{cuda}
{NVIDIA}.
\newblock {\em {NVIDIA} {CUDA} {C} Programming Guide (version 6.0)}, February
  2014.

\bibitem{owens-gpgpu}
J.D. Owens, M.~Houston, D.~Luebke, S.~Green, J.E. Stone, and J.C. Phillips.
\newblock {GPU} computing.
\newblock {\em Proceedings of the IEEE}, 96(5):879--899, May 2008.

\bibitem{streamscan}
Shengen Yan, Guoping Long, and Yunquan Zhang.
\newblock Streamscan: Fast scan algorithms for {GPU}s without global barrier
  synchronization.
\newblock In {\em Proceedings of the 18th ACM SIGPLAN Symposium on Principles
  and Practice of Parallel Programming}, PPoPP '13, pages 229--238, New York,
  NY, USA, 2013. ACM.

\end{thebibliography}

\end{document}